\documentclass[natbib]{iaus}
\usepackage{graphicx}

\title[Mass--Loss in LMC Cepheids] 
{Modeling Mass--Loss and Infrared Excess in Large Magellanic Cloud Cepheids}

\author[Neilson et al.]   
{Hilding R. Neilson$^1$, Chow--Choong Ngeow$^2$, Shashi M. Kanbur$^3$
 \and John B. Lester$^4$}

\affiliation{$^1$Department of Astronomy \& Astrophysics, University of Toronto, \\ 50 St. George Street,
 Toronto, ON, Canada, M5S 3H4 \\ email: {\tt neilson@astro.utoronto.ca} \\[\affilskip]
 $^2$ University of Illinois Urbana--Champain \\[\affilskip]
 $^3$ State University of New York Oswego \\[\affilskip]
$^4$University of Toronto Mississauga}

\pubyear{2008}
\volume{256}  
\pagerange{}
\jname{The Magellanic System: Stars, Gas, and Galaxies}
\editors{Jacco Th. van Loon \& Joana M. Oliveira, eds.}
\begin{document}

\maketitle

\begin{abstract}
The purpose of this preliminary work is to determine if Large Magellanic Cloud (LMC) Cepheids have stellar winds.  If  a Cepheid undergoes mass loss then at some distance from the star, a fraction of the gas becomes dust, which causes an infrared excess.  Mass loss is tested using OGLE II optical observations and SAGE infrared observations for a sample of 488 Cepheids.  The resultant mass-loss rates range from $10^{-12}$ to $10^{-7}$ $M_\odot/yr$.  Using the mass--loss model we compute infrared stellar luminosities for the sample of Cepheids and compare predicted infrared PL relations with observed relations.  The predicted relations not only vary from the observed relations, implying mass loss plays a significant role, but also show evidence for non-linearity.  It is determined that mass loss is important for LMC Cepheids.
\end{abstract}

\firstsection 
\section{Introduction}

LMC Cepheids are important tools for testing stellar astrophysics and for calibrating the Period-Luminosity (PL) relation at various wavelengths.   Recently infrared PL relations have been determined for LMC Cepheids (Ngeow \& Kanbur 2008; Freedman et al. 2008).  These PL relations are more universal because the pulsation amplitude decreases at longer wavelengths and metallicity contributes less.  
      
      Infrared observations of Galactic Cepheids have found the existence of circumstellar shells (Kervella et al. 2006; M\'{e}rand et al. 2006; M\'{e}rand et al. 2007) and it has been suggested that the shells are caused by stellar winds with dust shells forming at large distance from the star (Neilson \& Lester 2008).  The dust causes the observed infrared excess.   Mass loss is also a potential solution to the Cepheid mass discrepency which is the difference between mass estimates from stellar evolution calculations and stellar pulsation models (Cox 1980).  If LMC Cepheids have significant stellar winds then mass loss can explain the discrepancy and may also cause infrared excess.

\begin{figure}[t]
\begin{center}
 \includegraphics[width=2.6in]{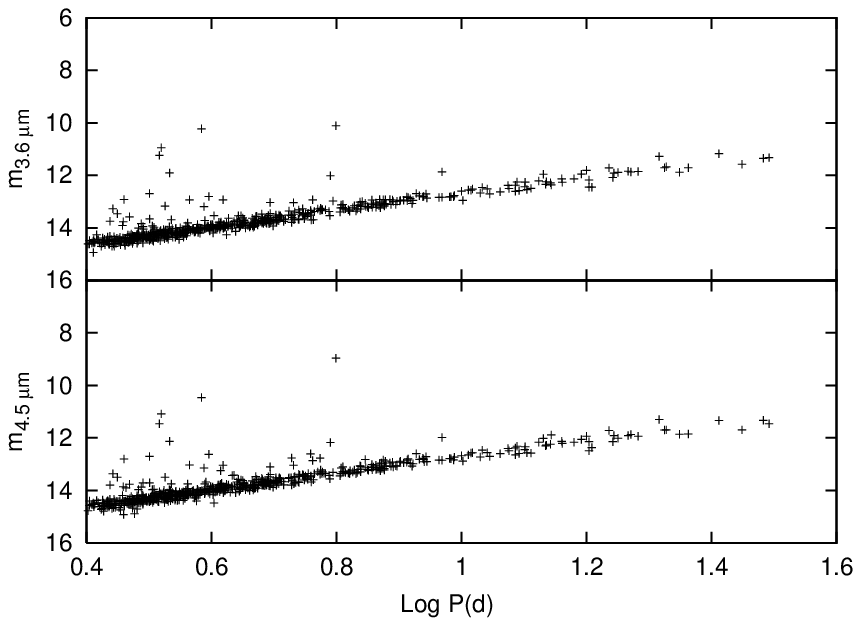}  \includegraphics[width=2.6in]{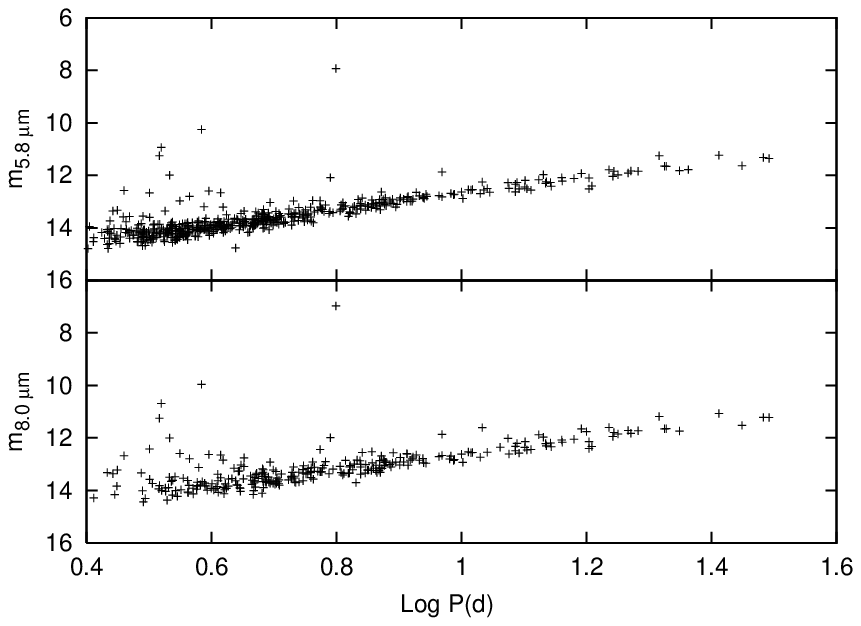} 
\end{center}
 \caption{The apparent magnitudes LMC Cepheids at infrared wavelengths.}
   \label{fig1}

\end{figure}

\section{Method}

We wish to determine mass-loss rates for LMC Cepheids.  To do this we use BV and I observations from the OGLE II survey and IRAC observations from the SAGE survey.  Having up to seven fluxes as a function of wavelength, we can $\chi^2$ fit mass-loss rates and radii by predicting stellar luminosities and circumstellar shell luminosities, which are
\begin{equation}
L_{\rm{Star},\nu} = 4\pi R^2_*\pi B_\nu(T_{\rm{eff}}),
\end{equation}
\begin{equation}
L_{\rm{Shell},\nu} = \frac{3}{4\pi}\frac{<a^2>}{<a^3>}\frac{1}{\bar{\rho}}\frac{\dot{M}_d}{v_d}Q^A_\nu\int_{R_*}^\infty B_\nu(T_{\rm{eff}})[1 - W(r)]dr.
\end{equation}
The effective temperature of a Cepheid is determined using the Temperature--Period--Color relation from Beaulieu et al. (2001).  The dust grains are assumed to have size $a$ that ranges from $0.005$ to $0.025$ $\mu m$, and the density of a grain is about $1 g/cm^3$, and $Q^A_\nu$ is the absorption efficiency.  The mass-loss rate is that of the dust and dust velocity is about the escape velocity of a Cepheid, $100 km/s$. The quantity $W(r)$ is the dilution factor and is a function of stellar radius.  The dust temperature scales as the effective temperature.  Therefore given a dust mass-loss rate and radius we can predict the total luminosity (stellar + shell) of a Cepheid at various frequencies and compare these to observations.  The gas mass-loss rate is assumed to be $250$ times larger than the dust mass-loss rate but this choice of dust-to-gas ratio leads to a minimum prediction of gas mass-loss rate.  We apply this analysis to a sample of 488 Cepheids with observed infrared fluxes shown in Figure 1 (Left).  

\section{Results}

The gas mass-loss rates  and best--fit $\chi^2$ values are shown in Figure 2.  The values imply that mass loss is important for LMC Cepheids and contribute to the observed infrared fluxes.  We take the predicted stellar fluxes and compute infrared PL relations to understand the contribution of mass loss on the observed relations.  The relations are shown in Figure 3 (Left) and the slopes and zero points are given in Table 1.
\begin{table}[t]
\caption{Best Fit Parameters for Predicted PL Relations}
\begin{center}
\begin{tabular}{lcccc}
\hline
Type & $\lambda$ $(\mu m)$ & Slope & Zero Point & Dispersion \\
\hline
Linear & $3.6$ & $-3.14 \pm 0.024$ & $15.99 \pm 0.017$ & $0.110$ \\
 &$4.5$ & $-3.15 \pm 0.023$ & $15.91 \pm 0.017$ & $0.108$\\
&$5.8$ & $-3.16 \pm 0.023$ & $15.92\pm 0.017$ & $0.107$ \\
& $8.0$ & $-3.17 \pm 0.022$ & $15.92 \pm 0.016$ & $0.105$ \\
\hline
Non--  & $3.6$ &$-3.24 \pm 0.038$ & $16.04 \pm 0.025$ & $0.107$ \\
 Linear & $4.5$ &  $-3.24 \pm 0.037$ & $15.97\pm 0.024$ & $0.106$ \\
$P <10d$ & $5.8$ & $-3.25 \pm 0.037$ & $15.97 \pm 0.024$ & $0.104$ \\
 & $8.0$ & $-3.26 \pm 0.036$ & $15.97 \pm 0.024$ & $0.103$ \\
 \hline
 Non-- & $3.6$ & $-2.97 \pm 0.13$ & $15.81 \pm 0.15$ & $0.125$ \\
Linear  & $4.5$ & $-2.99 \pm 0.12$ & $15.75 \pm 0.14$ & $0.122 $\\
$P>10d$& $5.8$& $-3.00 \pm 0.12$ & $15.75 \pm 0.14$ & $0.120 $\\
 & $8.0$ & $-3.02 \pm 0.12$ & $15.76 \pm 0.14$ & $0.118$ \\
\hline
\end{tabular}
\end{center}
\label{t1}
\end{table}

\begin{figure}[t]
\begin{center}
 \includegraphics[width=2.6in]{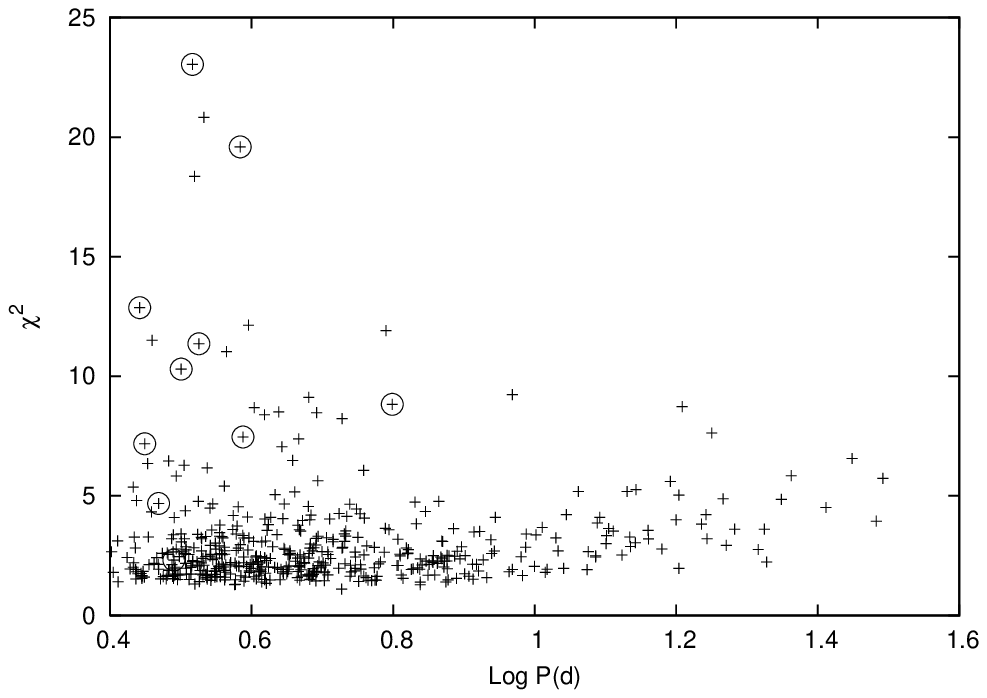}  \includegraphics[width=2.6in]{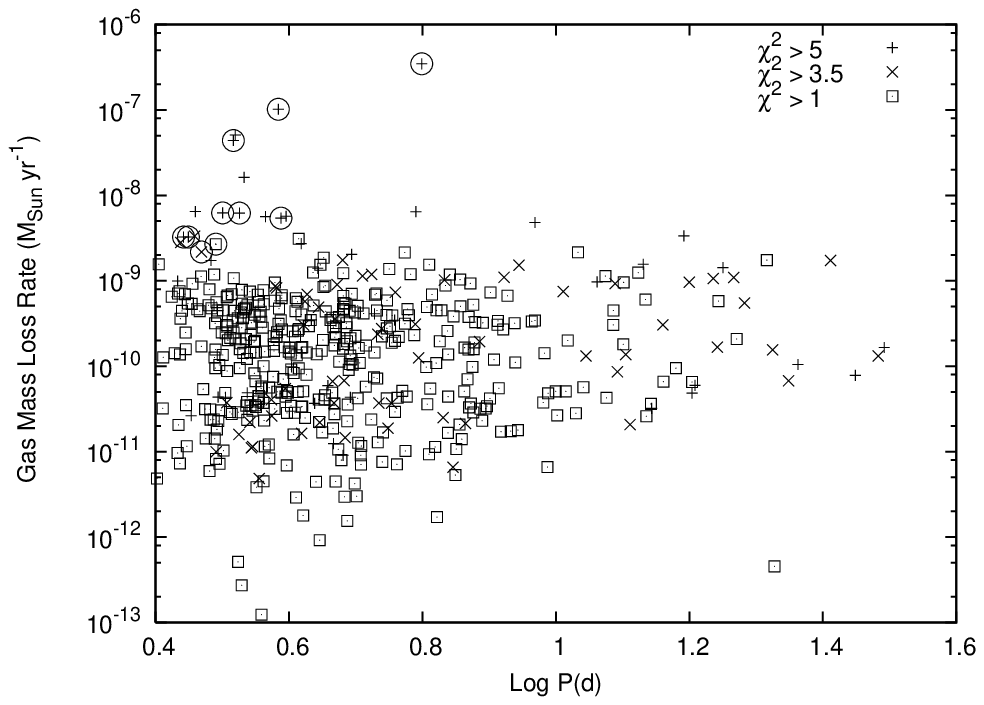}  
\end{center}
 \caption{(Left) The values of $\chi^2$ for the fits to the sample of Cepheids. (Right) The predicted mass--loss rates of the Cepheids. The circled points represent Cepheids where the infrared observations are likely blended stars. }
   \label{fig2}

\end{figure}

The infrared PL relations predicted here differ from those found in the literature (Ngeow \& Kanbur 2008;  Freedman et al. 2008) as can be seen in Figure 3 (Right).  The PL relations from \cite{Ngeow2008} tend to have brighter zero points, and similar slopes, resulting from larger infrared excesses from the shortest period Cepheids.  The predicted stellar IR PL relations have shallower slopes than the relations from \cite{Freedman2008} but similar zero points. 

\begin{figure}[t]
\begin{center}
 \includegraphics[width=2.6in]{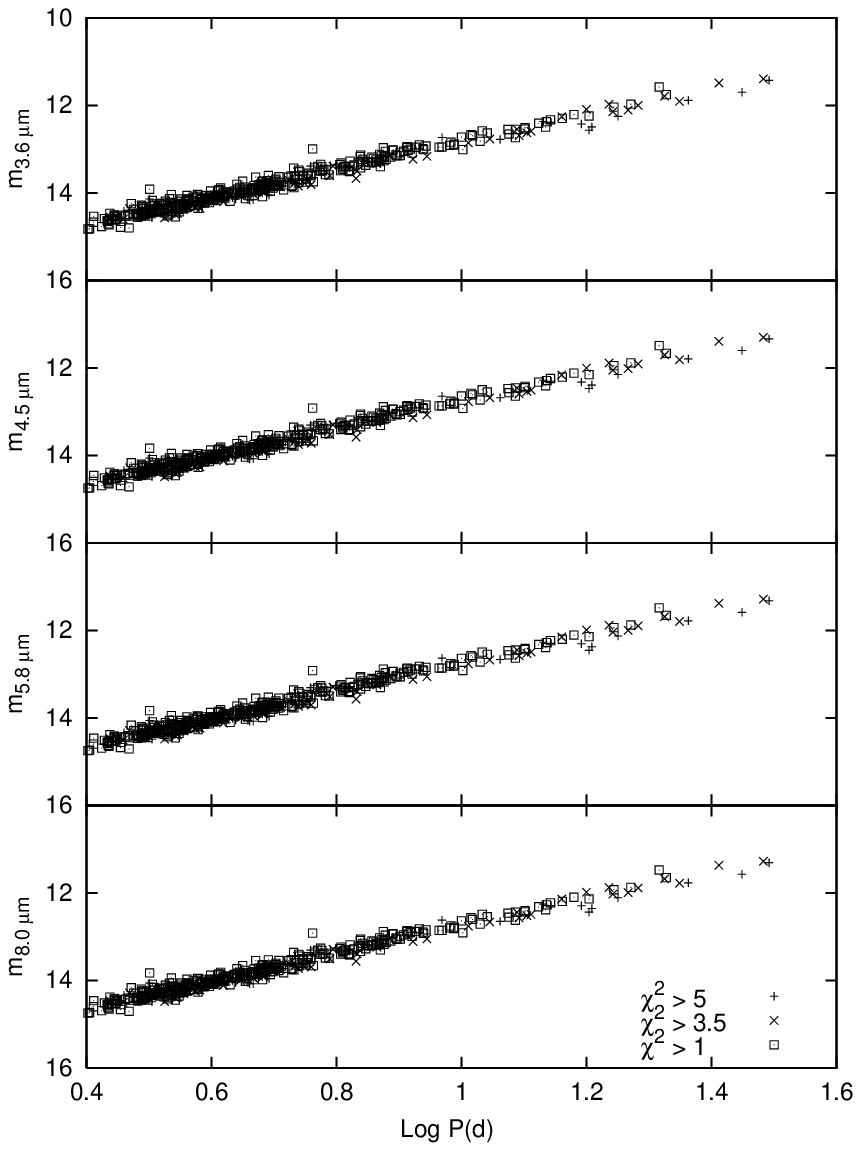}  \includegraphics[width=2.6in]{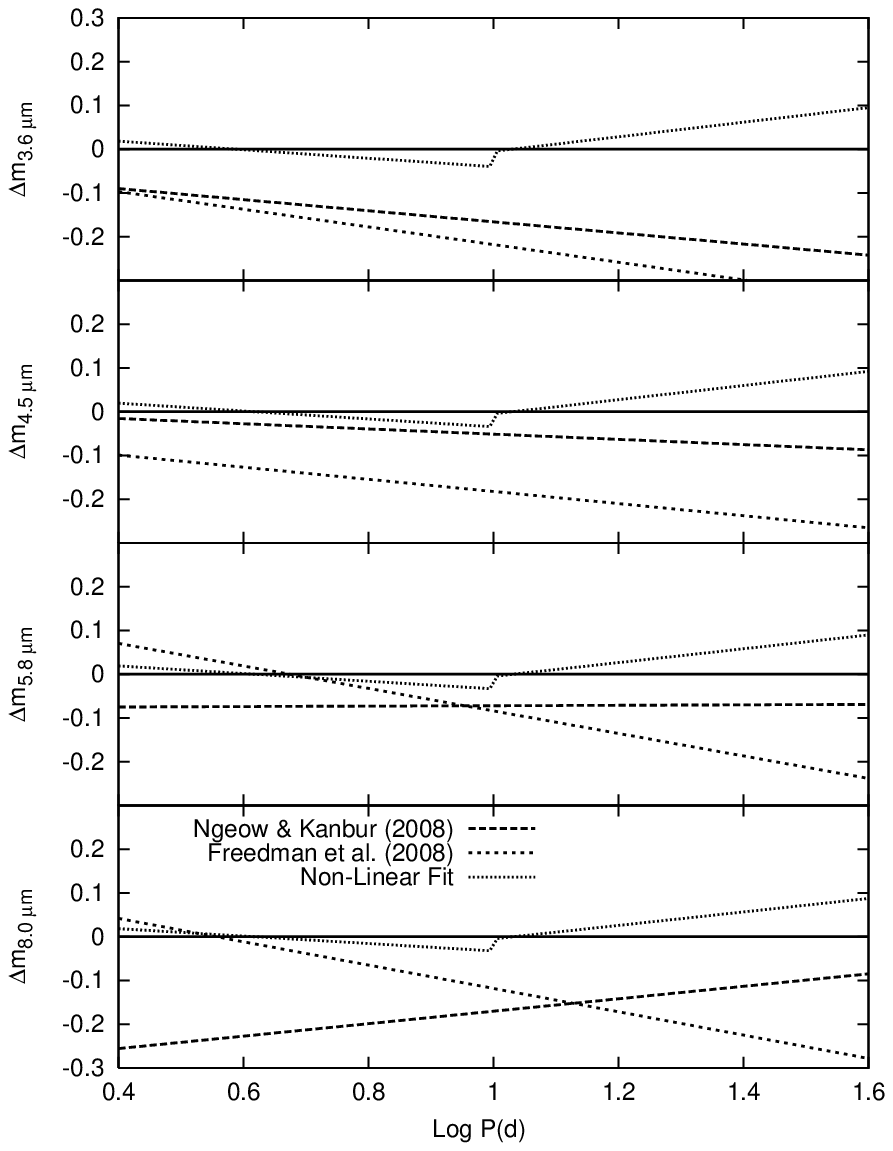} 
\end{center}
 \caption{(Left) The predicted apparent magnitudes of the sample of LMC Cepheids at infrared wavelengths. (Right) The comparison of the observed IR PL relations to our predicted linear PL relation.}
   \label{fig3}

\end{figure}

It has been shown that LMC PL relations are non-linear in the optical wavelengths (Ngeow et al. 2005) with the K-band relation being marginally linear (Ngeow \& Kanbur 2006).  F-tests of the \cite{Ngeow2008} found that the IR PL relations were linear from $3.6$ to $5.8$ $\mu m$ while the $8.0$ $\mu m$ relation is non-linear.  The optical non-linear PL relations are to have a period break at $10$ days, where the slope is steeper for $P < 10d$ than for $P > 10d$.  The $8.0$ $\mu m$ relationÕs slope is more shallow for the $P < 10d$.  We apply the F-test (Kanbur \& Ngeow 2004) to the predicted stellar fluxes and find that the predicted IR PL relations are consistent with being non-linear in a similar fashion as the optical PL relations wit the $8.0$ $\mu m$ data being marginally linear.  The slopes and zero points of the non-linear relations are given in Table 1 and the comparison of the predicted linear and predicted non-linear relations are shown in Figure 3 (Right).  It appears mass loss masks the non-linearity by causing a larger infrared excess in the shorter period Cepheids than in the long period Cepheids.  This is because the long period Cepheids are already redder and the mass-loss rates on average are similar for all periods.  This also explains the non-linear behavior of the observed $8.0$ $\mu m$ relation and explains why the K-band relation is marginally linear.

\section{Conclusions}

In this work, we have modeled mass loss in Cepheids as a spherically symmetric wind where dust grains form at some distance from the star where the temperature of the wind decreases to the condensation temperature.  The dust in the wind causes an infrared excess that can be observed.  By applying this model to OGLE BVI and SAGE IRAC observations we can determine mass-loss rates and analyze the role mass loss plays.  We find the following conclusions:
\begin{itemize}
\item Mass loss is significant in LMC Cepheids ranging from $10^{-12}$ to $10^{-7}$ $M_\odot/yr$.
\item The mass-loss rates imply that Cepheids may lose up to $1 M_\odot$ on the second crossing of the instability strip which is consistent with the mass discrepancy  (Keller 2008).  Therefore mass loss is a plausible solution.
\item Infrared excess due to mass loss have a significant contribution to the IR PL relations affecting the slope and zero point.
\item The predicted stellar IR PL relations show evidence for non-linearity.
\item The computed mass-loss rates are dependent on the dust-to-gas ratio which is metallicity dependent.  Because mass loss affects the structure of IR PL relation then the observed IR PL relations are metallicity dependent.
\end{itemize}


\begin{thebibliography}{}

\bibitem[Ngeow \& Kanbur (2008)]{Ngeow2008}
{Ngeow, C., \& Kanbur, S.} 2008,
\textit{ApJ}, 679, 76

\bibitem[Freedman et al. (2008)]{Freedman2008}
{Freedman, W.L., Madore, B.F., Rigby, J., Persson, S.E. \& Sturch, L.} 2008, 
\textit{ApJ}, 679, 71

\bibitem[Kervella et al. (2006)]{Kervella2006}
{Kervella, P., M\'{e}rand, A., Perrin, G., \& Coud\'{e} Du Foresto, V.} 2003, 
\textit{A\&A}, 448, 623

\bibitem[M\'{e}rand et al. (2006)]{Merand2006}
{M\'{e}rand, A., Kervella, P., Coud\'{e} Du Foresto, V., Perrin, G., Ridgway, S.T., Aufdenberg, J.P., Ten Brummelaar, T.A., McAlister, H.A., Sturmann, L., Sturmann, J., Turner, N.H., \& Berger, D.H} 2006, 
\textit{A\&A}, 453, 155

\bibitem[M\'{e}rand et al. (2007)]{Merand2007}
{M\'{e}rand, A., Aufdenberg, J.P., Kervella, P., Coud\'{e} Du Foresto, V., Ten Brummelaar, T.A., McAlister, H.A., Sturmann, L., Sturmann, J., \& Turner, N.H.} 2007, 
\textit{ApJ}, 664, 1093

\bibitem[Beaulieu et al. (2001)]{Beaulieu2001}
{Beaulieu, J.P., Buchler, J.R., \& Koll\'{a}th, Z.} 2001,
\textit{A\&A}, 373, 164

\bibitem[Neilson\& Lester (2008)]{Neilson2008}
{Neilson, H.R., \& Lester, J.B.} 2008,
\textit{ArXiv e--prints}, 803

\bibitem[Cox (1980)]{Cox1980}
{Cox, A.N. } 1980, 
\textit{ARA\&A}, 18, 15

\bibitem[Meixner, M., et al. (2006)]{Meixner2006}
{Meixner, M., Gordon, K.D., Indebetouw, R., et al.} 2006, 
\textit{AJ}, 132, 2268

\bibitem[Ngeow et al. (2005)]{Ngeow2005}
{Ngeow, C., Kanbur, S.M., Nikolaev, S., Buonaccorsi, J., Cook, K., \& Welch, D.} 2005, 
\textit{MNRAS}, 363, 831

\bibitem[Ngeow \& Kanbur (2006)]{Ngeow2006}
{Ngeow, C., \& Kanbur, S.M.} 2006, 
\textit{ApJ}, 650, 180

\bibitem[Kanbur \& Ngeow (2004)]{Kanbur2004}
{Kanbur, S.M., \& Ngeow, C.~C.} 2004, 
\textit{MNRAS}, 350, 962

\bibitem[Keller (2008)]{Keller2008}
{Keller, S.} 2008,
\textit{ApJ}, 677, 483
\end{thebibliography}
\end{document}